\documentclass[aps,pre,preprint,
groupedaddress,
]{revtex4-1}
\usepackage[textsize=small]{todonotes}
\usepackage{color}              
\usepackage{epsfig}
\usepackage{graphicx}
\usepackage{amsmath}
\usepackage{amsfonts}
\usepackage{amsthm}
\usepackage{amssymb}
\usepackage{epsfig}
\usepackage{hyperref}
\usepackage{bbm}
\usepackage{longtable}
\usepackage{tikz,pgfplots}
\usepackage{scalefnt}
\usetikzlibrary{plotmarks}

\definecolor{darkgreen}{rgb}{0,.4,0}
\definecolor{darkagenta}{rgb}{.5,0,.5}
\definecolor{darkred}{rgb}{1,0,0}
\definecolor{darkblue}{rgb}{0,0,.4}

\renewcommand{\P}{\mathbb{P}}

\renewcommand{\P}{\mathbb{P}}

\def\P{{\mathbb P}}

\def\v{{v}}

\newcommand{\eqn}[1]{\begin{equation} #1 \end{equation}}

\newcommand{\sss}{\scriptscriptstyle}

\newcommand{\expec}{{\mathbb E}}

\newcommand {\convd}{\stackrel{d}{\longrightarrow}}
\newcommand {\convp}{\stackrel{\sss {\mathbb P}}{\longrightarrow}}

\numberwithin{equation}{section}

\begin{document}

\title{{Uncovering disassortativity in large scale-free networks}}


\author{Nelly Litvak}
\email[]{n.litvak@ewi.utwente.nl}

\affiliation{University of Twente}

\author{Remco van der Hofstad}
\email[]{r.w.v.d.hofstad@TUE.nl}

\affiliation{Eindhoven University of Technology}

\date{\today}

\begin{abstract}

Mixing patterns in large self-organizing networks, such as the Internet, the World Wide
Web, social and biological networks are often characterized by degree-degree {dependencies} between neighbouring nodes.  In this paper we propose a new way of measuring degree-degree {dependencies}. One of the problems with the commonly used assortativity coefficient is that {in disassortative networks its magnitude decreases} with the network size. We mathematically explain this phenomenon and validate the results on synthetic graphs and real-world network data. As an alternative, we suggest to use rank correlation measures such as Spearman's rho. Our experiments convincingly show that Spearman's rho produces consistent values in graphs of different sizes but similar structure, and it is able to reveal strong (positive or negative) {dependencies} in large graphs. In particular, we discover much stronger negative {degree-degree dependencies} in Web graphs than was previously thought. {Rank correlations allow us to compare the assortativity of networks of different sizes, which is impossible with the assortativity coefficient due to its genuine dependence on the network size.} We conclude that rank correlations provide a suitable and informative method for uncovering network mixing patterns. 

\end{abstract}


\pacs{89.75.Hc, 87.23.Ge, 89.20.Hh}

\maketitle

\section{Introduction}
\label{sec:intro}

This paper proposes a new way of measuring mixing patterns in large self-organizing networks, such as the Internet, the World Wide Web, social and biological networks. 
Most of these real-world networks are scale-free, i.e., their degree distribution has huge variability and closely follows a power law (the fraction of nodes with degree $k$ is roughly proportional to $k^{-\gamma-1}$, $\gamma>0$). We study correlations between degrees of two nodes connected by an edge. This problem, first posed in~\cite{Newman2002assortative,Newman2003mixing}, has received vast attention in the networks literature, in particular in physics, sociology, {biology} and computer science. We show however, analytically and on the data, that the presence of power laws  makes currently used measures inadequate {for comparison of mixing patterns in networks of different sizes, and provide an alternative that is free from this disadvantage}.

{Adequate measuring and comparison of degree-degree correlations is important because mixing patterns define many  of the network's properties.} For instance, the Internet topology is not sufficiently specified by the degree distribution; the negative degree-degree correlations in the Internet graph have a great influence on the robustness to failures~\cite{Doyle2005ryf}, efficiency of Internet protocols~\cite{Li2005willinger}, {as well as} distances and betweenness~\cite{Mahadevan2006Internet_topology}. This is totally different from the mixing patterns in {networks} of bank transactions~\cite{May2008banks} where the core of 25 most important banks is entirely connected. The correlation between in- and out-degree of tasks plays and important role in the  dynamics of production and development systems~\cite{Braha2007communication}. Mixing patterns affect epidemic spread~\cite{Eguiluz2002epidemic,Eubank2004nature} and Web ranking~\cite{Fortunato2007mean_field}. 

In his seminal papers, Newman~\cite{Newman2002assortative,Newman2003mixing} proposed to measure {degree-degree correlations} using {the} {\it assortativity} coefficient, which is, in fact, an empirical {estimate} of the {Pearson's} correlation coefficient between the degrees {at either ends} of a random edge.  {A} network is \emph{assortative} when neighbouring nodes {are likely to} have a similar number of connections. In \emph{disassortative} networks, high-degree nodes mostly have neighbours with small number of connections.
The empirical data in \cite[Table I]{Newman2002assortative} suggest that
social networks tend to be
assortative (which is indicated by the positive assortativity coefficient), while technological and biological networks tend to be disassortative.

In \cite[Table I]{Newman2002assortative}, it is striking that larger 
disassortative networks {typically} have an 
assortativity coefficient that is closer to 0 and therefore
appear to have {approximately} {\it uncorrelated} degrees
across edges. Similar conclusions can be drawn from \cite[Table II]{Newman2003mixing}. 
{In recent literature ~\cite{Dorogovtsev2010correlationsPA,Raschke2010correlations} the issue was raised  that the Pearson's correlation coefficient in scale-free networks {decreases with} the network size.} In this paper we demonstrate analytically and on the data that in {\it all} scale-free disassortative networks with {a realistic value of the power-law exponent}, the assortativity  coefficient {decreases in magnitude with the size of the graph}. In assortative networks, on the other hand, the assortativity coefficient can show two types of behaviour. It either decreases with graph size, or it shows a considerable dispersion in values, even if large networks are constructed by the same mechanism. 

We suggest an alternative solution based on the classical Spearman's rho measure~\cite{Spearman1904} that {is the correlation coefficient computed on the \emph{ranks}} of degrees. The huge advantage of such dependency measures is that they
work well \emph{independently} of the degree distribution,  while the assortativity coefficient, despite the fact that it is 
always in $[-1,1]$, suffers from a strong dependence on the extreme values of
the degrees. The {usefullness of the rank correlation approach to discover dependencies in skewed distributions has already been postulated} in the 1936 paper by H.~Hotelling and M.R.~Pabst~\cite{Hotelling1936}:
{\it `Certainly where there is complete absence of knowledge of the form of the bivariate distribution, and especially if it is believed not to be normal, the rank correlation coefficient is to be strongly recommended as a means of testing the existence of relationship.'}  

We compute Spearman's rho on artificially generated random graphs and on real data from web and social networks. Our results agree with~\cite{Newman2002assortative} concerning the presence of positive or negative correlations, but Spearman's {rho} has two important advantages: {(1)} it is able to reveal strong {disassortativity in large networks;} {(2)} it produces consistent values on the graphs created by the same mechanism, e.g.\ on {preferential attachment} graphs~\cite{Albert99} of different sizes. Thus, Spearman's rho correctly {and consistently} captures the underlying connection patterns and tendencies. {We conclude that when networks are large, or two networks of difference sizes must be compared (e.g.\ in web crawls or social networks from different countries), Spearman's rho is a preferred method for measuring and comparing degree-degree correlations.}

{The closing section discusses further challenges in the evaluation of network mixing patterns.}

\section{No disassortative scale-free random graph sequences}
\label{sec-no-disass-RG-seq}

In this section we present a simple analytical argument that in disassortative networks {the assortativity coefficient always decreases in magnitude with the size of the graph.} Formal proofs can be found in \cite{Litvak2012correlations}.

Assortativity in networks is usually measured using the assortativity coefficient, which is in fact a statistical estimator of a Pearson's correlation coefficient for the degrees on the two ends of an arbitrary edge in a graph. Let $G=(V,E)$ be a graph with vertex set $V$, {where $|V|=n$ denotes the size of the network,} and edge set $E$.
The assortativity coefficient of $G$ is equal to
(see, e.g., \cite[(4)]{Newman2002assortative})
\begin{equation}
    \label{eq:assortativity_def}
    \rho_n=\frac{\frac{1}{|E|}\sum_{ij\in E} d_id_j-\Big(\frac{1}{|E|} \sum_{ij\in E} \tfrac12 (d_i+d_j)\Big)^2}
    {\frac{1}{|E|}\sum_{ij\in E} \tfrac12(d_i^2+d_j^2) -\Big(\frac{1}{|E|} \sum_{ij\in E} \tfrac12 (d_i+d_j)\Big)^2},
    \end{equation}
where the sum is over directed edges of $G$,  i.e., $ij$ and $ji$ are two distinct edges, and $d_i$ is the degree of vertex $i$. We compute that 
    \[
    \frac{1}{|E|} \sum_{ij\in E} \tfrac12 (d_i+d_j)
    =\frac{1}{|E|} \sum_{i\in V} d_i^2,
    \qquad
    \frac{1}{|E|}\sum_{ij\in E} \tfrac12(d_i^2+d_j^2)
    =\frac{1}{|E|} \sum_{i\in V} d_i^3.
    \]
Thus, $\rho_n$ can be written as
    \eqn{
    \label{eq:rhoG}
    \rho_n=\frac{\sum_{ij\in E} d_id_j-\frac{1}{|E|} \Big(\sum_{i\in V} d_i^2\Big)^2}
    {\sum_{i\in V} d_i^3 -\frac{1}{|E|} \Big(\sum_{i\in V} d_i^2\Big)^2}.
    } 

In practice, all quantities in (\ref{eq:rhoG}) are finite, and $\rho_n$ can always be computed. However, since many real-life networks are very large, a relevant question is {how $\rho_n$ behaves when $n$ becomes large.}

In the literature, many examples are reported of real-world networks
where the degree distribution obeys a power law~\cite{Albert2002stat_mech,Newman03survey}.
In particular, for scale-free networks, the observed proportion of vertices of degree 
$k$ is close to $f(k)=c_0 k^{-\gamma-1}$, and {most values of $\gamma$ found in real-world networks} 
are in $(1,3)$, see e.g., \cite[Table I]{Albert2002stat_mech}
or \cite[Table I]{Newman03survey}. For $p<\gamma$, let $\mu_p=\sum_kk^pf(k)$, and note that the series diverges if $p\ge \gamma$; let $a\sim b$ denote that $a/b\to 1$. Then we
can expect that, as $n$ grows large,
    \[
    |E|= \sum_{i\in V} d_i\sim \mu_1 n,\quad \sum_{i\in V} d_i^p\sim  \mu_p n,\quad p<\gamma,
    \]
while $\max_{i\in V}d_i$ is of the order $n^{1/\gamma}$. As a direct consequence,
	\begin{align}
	\label{eq:cond1}
	& cn\leq |E|\leq Cn,\\
	\label{eq:cond2}
	&c n^{1/\gamma}\le \max_{i\in [n]} d_i \leq C n^{1/\gamma},\\
	\label{eq:cond3}
	&cn^{\max\{p/\gamma,1\}}\leq \sum_{i\in [n]} d_i^p \leq Cn^{\max\{p/\gamma,1\}}, \quad p=2,3,
	\end{align}
for  $\gamma\in (1,3)$ and some constants $0<c<C<\infty$. We emphasize  that conditions (\ref{eq:cond1}) -- (\ref{eq:cond3}) are very general and hold for any scale-free network {of growing size}, independently of its mixing patterns. 
{From} (\ref{eq:rhoG}) we simply write
    \[
    \rho_n\geq \rho_n^-\equiv 
    -\frac{\frac{1}{|E|}\Big(\sum_{i\in V} d_i^2\Big)^2}
    {\sum_{i\in V} d_i^3-\frac{1}{|E|}\Big(\sum_{i\in V} d_i^2\Big)^2},
    \]
and notice that \[\sum_{i\in V} d_i^3\geq (\max_{i\in [n]} d_i)^3
\geq c^3 n^{3/\gamma},\] whereas 
\[\frac{1}{|E|}\Big(\sum_{i\in V} d_i^2\Big)^2
\leq (C^2/c) n^{2\max\{2/\gamma,1\}-1}=(C^2/c) n^{\max\{4/\gamma-1,1\}}.\] Since $\gamma\in (1,3)$ we have $\max\{4/\gamma-1,1\}<3/\gamma$, so that
	\[
	\frac{\sum_{i\in V} d_i^3}{\frac{1}{|E|}\Big(\sum_{i\in V} d_i^2\Big)^2}
	\rightarrow \infty\quad \mbox{as $t\to\infty$}.
	\]
Hence, the lower bound ${\rho}^-_n$ 
is of the order $n^{\max\{1/\gamma-1,1-3/\gamma\}}$. It is now easy to check that if $\gamma\in (1,3)$, then
$\rho_n^-$ converges to zero when the graph size increases. This means that any limit point of the assortativity coefficients $\rho_n$ is non-negative. {Note also that $\rho_n^-$ is defined by the degree sequence, and it does not depend on the mixing pattern at all. We conclude that by looking only at the value of $\rho_n$ one cannot  discover even very strong disassortativity in large scale-free graphs.} We will confirm this finding in Section~\ref{sec:RG} on artificially generated random graphs, and in Section~\ref{sec:num} {on} real-world networks.

{We note that if $\gamma>3$, then all terms in (\ref{eq:assortativity_def}) converge to a number, and $\rho_n$ does not scale with the network size. In practice this means that the dependence of $\rho_n$ on the graph size is observed when node degrees have a broad distribution, and this range increases when the network gets bigger. This is the case in most real-life networks and models for them, as is e.g.\ obviously the case for preferential attachment models.}

We {further notice} that (\ref{eq:cond1})--(\ref{eq:cond3}) imply that
	\begin{align}
	\label{eq:crossproducts1}
	&\sum_{ij\in E} d_id_j
	\le \Big(\max_{i\in[n]}d_i\Big)\sum_{ij\in E_n} d_i
	=\max_{i\in[n]}d_i\Big(\sum_{i\in V_n} d^2_i\Big)
	\le C^2n^{1/\gamma+\max\{2/\gamma,1\}}.
	\end{align}
Mathematically, an interesting 
case is when $\sum_{ij\in E} d_id_j$ and ${\sum_{i\in V} d_i^3}$ are of the same order of magnitude. {Then} the network {is} assortative but, formally, ${\rho_n}$ converges to a random variable. In practice this means that $\rho_n$ can result in very different values on two very large graphs constructed by the same mechanism. We will give such an {example} in Section~\ref{sec:RG}. 

\section{Rank correlations}
\label{sec:spearman}

We propose an alternative measure for the degree-degree dependencies, based on the rank correlations. For two-dimensional data $((X_i, Y_i))_{i=1}^n$, let $r_i^{X}$ and $r_i^{Y}$ be the rank of an observation $X_i$ and $Y_i$, respectively,  when the sample values $(X_i)_{i=1}^n$ and
$(Y_i)_{i=1}^n$ are arranged in a descending order. The rank correlation measures evaluate statistical dependences on the data $((r_i^X,r_i^Y))_{i=1}^n$, rather than on the original data $((X_i, Y_i))_{i=1}^n$. 
Rank transformation is convenient, in particular because $(r^X_i)$  and $(r_i^Y)$ are samples from the same uniform distribution, {which} implies many nice mathematical properties. 

The statistical correlation coefficient for the rank is known as Spearman's rho~\cite{Spearman1904}:
\begin{equation}
\label{eq:spearman}
\rho_n^{\rm rank}=\frac{\sum_{i=1}^n(r_i^X-(n+1)/2)(r_i^Y-(n+1)/2)}{\sqrt{\sum_{i=1}^n(r_i^X-(n+1)/2)^2\sum_i^n(r_i^Y-(n+1)/2)^2}}.
\end{equation}
The mathematical properties of the Spearman's rho {have been extensively investigated}. In particular, if $((X_i, Y_i))_{i=1}^n$ consists of independent realizations of $(X,Y)$, and the joint distribution function of $X$ and $Y$ is differentiable, then $\rho_n^{\rm rank}$ is a consistent statistical estimator, and its standard deviation is of the order $1/\sqrt{n}$ independently {of} the exact form of the underlying distributions, see e.g. \cite{Borkowf2002}.

For a graph $G$ of size $n$, we propose to compute $\rho_n^{\rm rank}$ using (\ref{eq:spearman}) as follows. We define the random variables $X$ and $Y$ as the degrees on two ends of a random {\it undirected} edge in a graph (that is, when rank correlations are computed, $ij$ and $ji$ represent the same edge). For each edge, when the observed degrees are $a$ and $b$, we assign $[X=a,Y=b]$ or $[X=b,Y=a]$ with probability $1/2$. {Many values of $X$ and $Y$ will be the same making their rank ambiguous.} We resolve this by we adding independent {uniformly distributed random variables on $[0,1]$} to each value of $X$ and $Y$. {In the setting when the realisations $(X_i,Y_i)$ are independent, this way of resolving ties preserves the original value of the Spearman's rho on the population, see {e.g.~\cite{Mesfioui2005Spearman}. We refer to \cite{Nevslehova2007rank_correlations}} for a general treatment of rank correlations for non-continuous distributions.}

In the remainder of the paper we will demonstrate that {the measure $\rho_n^{\rm rank}$ gives consistent results for different $n$, and it is able to reveal strong negative degree-degree correlations in large networks.}

\section{Random graph data}
\label{sec:RG}

We consider four random graph models to highlight our results.

\begin{figure}
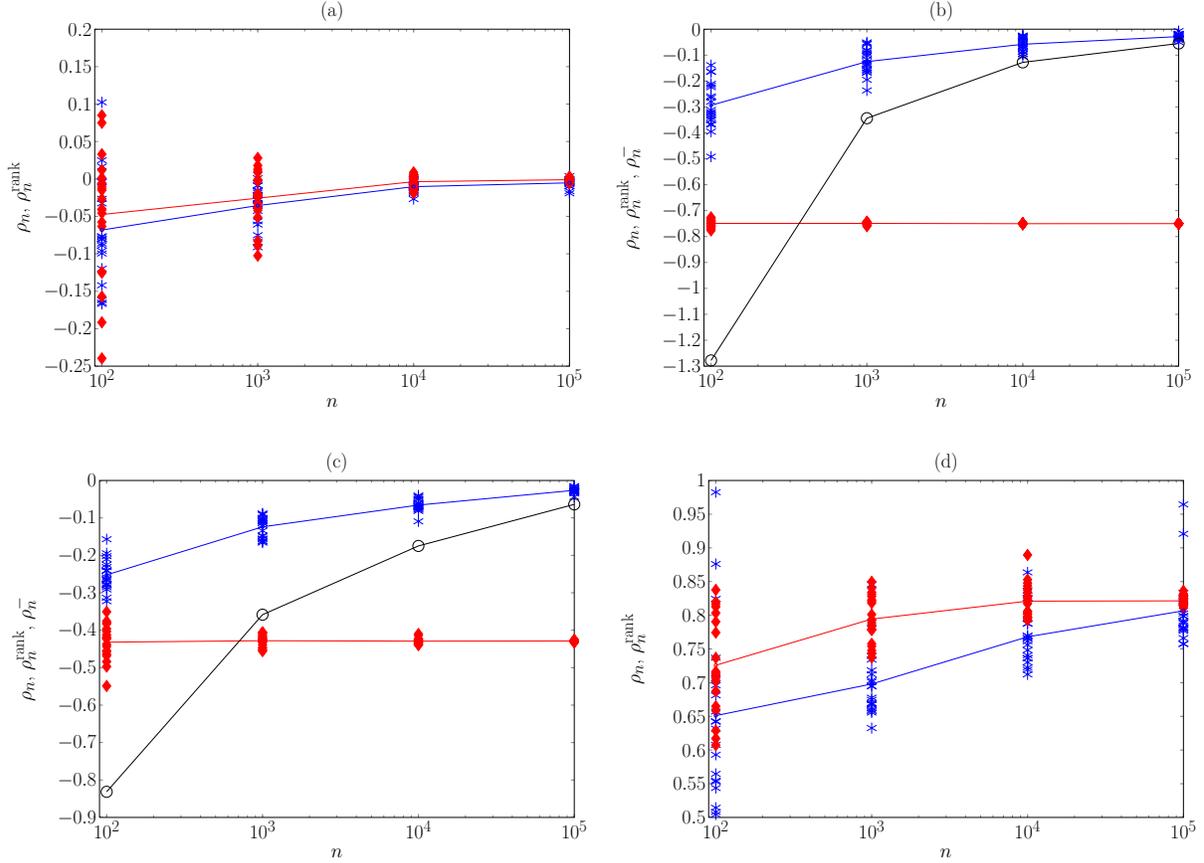
%
\include{Figure1}
 %
\caption{ (Color online) Scatter plots for samples of 20 graphs. For each size we plot the 20 realizations of $\rho_n$ (blue asterisks) and $\rho_n^{\rm rank}$ (red diamonds) in random graphs of different sizes. Solid lines connect the averages of the samples. In (c), (d) the circles connected by the solid line are the averages of $\rho_n^-$ in the samples. (a) Configuration model, $\P(d\ge x)=x^{-2}$, $x\ge 1$. (b) Configuration model with intermediate vertices.  (c) Preferential attachment model. (d) A collection of bi-partite graphs, where $b=1/2$, $a=2$, and $U$ has a generalized Pareto distribution $\P(U>x)=((2.1+x)/3.1)^{-3.1}$, $x> 1$.}%
\label{fig:RG}%
\end{figure}

\paragraph{The configuration model.}
The \emph{configuration model} was invented by 
Bollob\'as in \cite{Bollobas1980CM}, inspired by
\cite{Bender1978CM}. 
It was popularized by Newman, {Strogatz} and Watts~\cite{Newman2001CM}, who realized
that it is a useful and simple model for real-world networks. In the configurations model a node $i$ has a given number $d_i$ of {half-edges,} with $\ell_n=\sum_{i\in V} d_i$ assumed to be even. Each {half-edge} is connected to a randomly chosen {other half-edge to form an edge in the graph.}  {We chose $\gamma=2$, thus, the maximum degree is of the order $n^{1/2}$, which corresponds to the case of uncorrelated random networks, such that the probability that two vertices are directly connected is close to $d_id_j/\ell_n$~\cite{Boguna2004finite-size-effects,Catanzaro2005uncorrelated_networks}. Although self-loops and multiple edges can occur these become rare as $n\to\infty$, see e.g. \cite{Bollobas_RG} or \cite{Janson2009CM}.}
In simulations, we collapse multiple edges to a single edge, and remove self-loops. This changes the degree distribution slightly, and intuitively should yield negative dependencies. In Figure~\ref{fig:RG}(a) we observe that, on average, $\rho_n$ and $\rho_n^{\rm rank}$ are indeed negative in smaller networks {but then they converge to zero showing that the degrees on two ends of a random edge are uncorrelated.}

\paragraph{Configuration model with intermediate vertices.} {In order to construct a strongly disassortative graph, we first generate a configuration model as described above, and then we replace every edge by two edges that meet at a middle vertex.} In this model, 
there are $n+\ell_n/2$ vertices and $2\ell_n$ edges (recall that $ij$ and $ji$ are two different edges).
Now, if $E$, $V$, and $d_i$, $i=1,\ldots,n$ denote, respectively, the edge set, the vertex set, and the degrees of the original configuration model, then in the model with intermediate edges the assortativity coefficient is as follows:
	\[\rho_n=
	\frac{2\sum_{i\in V} 2d_i-\frac{1}{2\ell_n} \Big(\sum_{i\in V} d_i^2{+}2\ell_n\Big)^2}
    	{\sum_{i\in V} d_i^3 + 4\ell_n -\frac{1}{2\ell_n} \Big(\sum_{i\in V} d_i^2{+}2\ell_n\Big)^2}.
    	\]
{When} $\gamma<3$ we have $\mu_3=\infty$, and thus $\rho_n\to 0$ as $n\to\infty$. Furthermore, the {lower} bound $\rho_n^-$ also converges to zero as $n$ grows. It is clear that this particular random graph, of any size, is {equally} and strongly disassortative, however, $\rho_n$ fails to {capture this}. In Figure~\ref{fig:RG}(b) it is clearly seen that both $\rho_n$ and $\rho_n^-$ {quickly decrease in magnitude as $n$ grows}. It is striking that $\rho_n^{\rm rank}$ shows a totally different and very appropriate behavior. Its values remain around $-0.75$ identifying the strong negative dependencies, and the dispersion across different realizations of the graph decreases as $n\to\infty$.

\paragraph{Preferential attachment model.} We consider the basic version of the undirected preferential attachment model (PAM), where each new vertex adds only one edge to the network, connecting to the existing nodes with probability proportional to their degrees~\cite{Albert99}.  In this case, it is well known that $\gamma=2$
(see e.g. \cite{Bollobas01}). Newman~\cite{Newman2002assortative} noticed the counterintuitive fact that the Preferential Attachment graph has asymptotically neutral mixing, $\rho_n\to 0$ as $n\to\infty$. {This phenomenon has been studied in detail by Dorogovtsev et al.~\cite{Dorogovtsev2010correlationsPA}, and it can be clearly observed in Figure~\ref{fig:RG}(c).} The reason for this behavior is not the genuine neutral mixing in the {PAM} but rather the unnatural dependence of $\rho_n$ on the graph size. Indeed, we see that {PAMs of small sizes} have $\rho_n<0$, and then the magnitude of $\rho_n$ decreases with the graph size. Again, Spearman's {rho} consistently shows that the degrees are negatively
dependent. This can be understood by noting that the majority of edges of vertices
with high degrees, which are old vertices,  come from vertices which are added
late in the graph growth process and thus have small degree. On the other hand,
by the growth mechanism of the PAM, vertices with low degree are
more likely to be connected to vertices having high degree, which indeed suggests
negative degree-degree dependencies. 

\paragraph{A collection of complete bipartite graphs.} 
{We next present an example where the assortativity coefficient has a nonvanishing dispersion.} 
Take $((X_i,Y_i))_{i=1}^n$ to be a sample of independent realizations of the vector $(X,Y)$. We assume that $X={b}U_1+bU_2$ and $Y=bU_1+{a}U_2$, where $b>0$, $a>1$, and $U_1, U_2$ are independent identically distributed (i.i.d.) random variables with power law tail, and tail exponent $\gamma$.
Then, for $i=1, \ldots, n$, we create a complete bipartite graph of 
$X_i$ and $Y_i$ vertices, respectively. These $n$ complete bipartite graphs
are not connected to one another. We denote such {a} collection of $n$ bipartite 
graphs by $G_n$. This is an extreme scenario of a network consisting of highly connected clusters of different size. Such networks can serve as models for physical human contacts and are used in epidemic modelling~\cite{Eubank2004nature}.

The graph $G_n$  has
$|V|=\sum_{i=1}^n (X_i+Y_i)$ vertices and $|E|=2\sum_{i=1}^n X_iY_i$ edges.  
Further,
    	\[
    	\sum_{i\in V} d_i^p
    	=\sum_{i=1}^n (X_i^pY_i+Y_i^pX_i),
	\qquad
	\sum_{ij\in E}d_id_j=2\sum_{i=1}^n (X_iY_i)^2.
    	\]
Assume that $\P(U_j>x)= c_0 x^{-\gamma}$, where $c_0>0$, $x\ge x_0$, and $\gamma\in(3,4)$, so that $\expec[U^3]<\infty$,
but $\expec[U^4]=\infty$. As a result, $|E|/n \convp 2\expec[XY]<\infty$
and $\frac{1}{n} \sum_{i\in V} d_i^2\convp \expec[XY(X+Y)]<\infty$. Further,
	\[
	n^{-4/\gamma}b^{-4}\sum_{i=1}^n (X_i^3Y_i+Y_i^3X_i)
	\convd (a^3+a) Z_1+ 2Z_2,
	\qquad
	n^{-4/\gamma}b^{-4}\sum_{i=1}^N (X_iY_i)^2
	\convd a^2 Z_1+Z_2,
	\]
where $Z_1$ and $Z_2$ and two independent stable distributions with parameter $\gamma/4$.
As a result,
	\[
	\rho_n\convd \frac{2a^2Z_1+2Z_2}{(a+a^3)Z_1+2Z_2},\qquad \mbox{as $n\to\infty$},
	\]
which is a proper random variable taking values in $(2a/(1+a^2), 1)$, 
see~\cite{Litvak2012correlations} for detailed proof. 

Note that in this model there is a genuine dependence between the correlation 
measure and the graph size. Indeed, if $n=1$ then the assortativity coefficient 
equals $-1$ because nodes with larger degrees are connected to nodes with 
smaller degrees. However, when the graph size grows, the positive linear dependence between $X$ and $Y$
starts dominating, thus, larger graphs of this structure are strongly assortative. {While the example we present is quite special, we believe
that the effect described is rather general.}
 
In Figure~\ref{fig:RG}(d) we {again} see that $\rho_n^{\rm rank}$ captures the relation faster and gives 
consistent results with decreasing dispersion. On a contrary, $\rho_n$ has a persistent dispersion in its values, and we know from the result above that this dispersion will not vanish as $n\to\infty$. In the limit, 
${\rho_n}$ has a non-zero density on $(0.8,1)$. However, the convergence is too slow to observe it at $n=100,000$, because the vanishing terms are of the order $n^{-1/\gamma}$, which is only $n^{-1/3.1}$ in our example.

\section{Web samples and social networks} 
\label{sec:num}

We computed $\rho_n$, $\rho_n^{\rm rank}$ and $\rho_n^-$ on several Web samples (disassortative networks) and social network samples (assortative networks). We used the compressed graph data from the Laboratory of Web Algorithms (LAW) at the Universit\`a degli studi di Milano~\cite{Boldi2004,Boldi2011}. We used the bvgraph MATLAB package~\cite{Gleich2010ComputingPageRank}. The {\it stanford-cs} database~\cite{Constantine2007} is a 2001 crawl that includes all pages in the cs.stanford.edu domain. In datasets (iv), (vii), (viii) we evaluate $\rho_n$, $\rho_n^{\rm rank}$ and $\rho_n^-$ over 1000 random edges, and present the average over 10 such evaluations (in 10 samples of 1000 edges, the observed dispersion of the results was small). 

The results are presented in Table~\ref{tab:data}.
\begin{table}{\footnotesize
\begin{tabular}{|c|l|l|c|c|c|c|c|c|}
\hline
nr&Dataset&Description&\# nodes& \# edges& max degree&$\rho_n$&$\rho^{\rm rank}_n$&$\rho^-_n$\\
\hline
(i)&stanford-cs&web domain&9,914&54,854&340&-.1656&-.1627&-.4648\\\hline
(ii)&eu-2005&.eu web domain&862,664&5,477,938&68,963&-.0562&-.2525&-.0670\\\hline
(iii)&uk@100,000&.uk web crawl&100,000&5,559,150&55,252&-.6536&-.5676&-1.117\\\hline
(iv)&uk@1,000,000&.uk web crawl&1,000,000&77,123,940&403,441&-.0831&-.5620&-.0854\\\hline
(v)&enron&e-mail exchange&69,244&506,898&1,634&-.1599&-.6827&-.1932\\\hline
(vi)&dblp-2010&co-authorship&326,186&1,615,400&238&.3018&.2604&-.7736\\\hline
(vii)&dblp-2011&co-authorship&986,324&6,707,236&979&.0842&.1351&-.2963\\\hline
(viii)&hollywood-2009&co-starring&1,139,905&113,891,327&11,468&.3446&.4689&-0.6737\\\hline
\end{tabular}
}
\caption{(i)--(iv) Web crawls: nodes are web pages, and an (undirected) edge means that there is a hyperlink from one of the two pages to another; (iii),(iv) are  breadth-first crawls around one page. (v) e-mail exchange by Enron employees (mostly part of the senior management): node are employees, and an edge means that an e-mail message was sent from one of the two employees to another. (vi), (vii) scientific collaboration networks extracted from the DBLP bibliography service: each vertex represents a scientist and an edge means a co-authorship of at least one article. (viii) vertices are actors, and two actors are connected by an edge if they appeared in the same movie.}
\label{tab:data}
\end{table}
We {clearly see} that {the} assortativity coefficient $\rho_n$ and Spearman's $\rho_n^{\rm rank}$ always agree about whether dependencies are positive or negative. They also agree in magnitude of correlations when graph size is small or the lower bound $\rho_n^-$ is sufficiently far from zero. However, $\rho_n$ is {not consistent for graphs of similar structure but different sizes.} This is especially apparent on the two .uk crawls (iii) and (iv). Here $\rho_n$ is significantly smaller in magnitude on a larger crawl. Intuitively, mixing patterns should not depend on the crawl size. This is indeed confirmed {by} the value of Spearman's {rho}, which consistently shows strong negative correlations in both crawls. We could not observe a similar phenomenon so sharply in (vi) and (vii), probably because a larger co-authorship network incorporates articles from different areas of science, and the culture of scientific collaborations can vary greatly from one research field to another. 

We also notice that, as predicted by our results, the assortativity {coefficient tends to take smaller values than $\rho_n^{rank}$ if $\rho_n^-$ is small in magnitude. } This is clearly seen in the data sets (ii), (iv) and (v). Again, (ii) and (iv) are the largest among the analyzed web crawls.

The observed behaviour of the assortativity coefficient is explained by the above stated results that $\rho_n$ is {influenced greatly by the large dispersion in the degree values}. The latter increases with graph size because of the scale-free phenomenon. As a result, $\rho_n$ {becomes smaller in magnitude, which makes it impossible to compare graphs of different sizes}. In contrast, the {\it ranks} of the degrees are drawn from a uniform distribution on $[0,1]$, scaled by the factor $n$. Clearly, when a correlation coefficient is computed, the scaling factor cancels, and therefore Spearman's rho provides consistent results in the graphs of different sizes.

\section{Discussion}
\label{sec:discussion}

{The assortativity coefficient $\rho_n$ proposed in \cite{Newman2002assortative,Newman2003mixing} has been the first dependency measure introduced to describe degree-degree correlations in networks.
The assortativity coefficient has provided many interesting insights. It has been successfully used for comparison of dependencies in graphs with the same degree sequences~\cite{Maslov2002rewiring,Maslov2004rewiring}, and to generate graphs with given degrees and desired mixing patterns~\cite{Mieghem2010assortativity_spectrum}. 
An important drawback of $\rho_n$ is its dependence on the network size $n$. It has been noticed by many authors, and shown in this paper for disassortative networks, that $\rho_n$ converges to zero as $n$ grows.
In particular, the decay with network size of the assortativity coefficient $\rho_n$ implies that it cannot be used for comparing dependencies in networks of different sizes. Therefore, it prohibits the investigation whether growing networks become more or less assortitative over time.

This paper suggests to use rank-correlation measures such as Spearman's rho. Our experiments convincingly show that Spearman's rho does not suffer from the size-dependence deficiency. In networks of different sizes but similar structure, Spearman's rho yields consistent results, and it is able to reveal strong (positive or negative) correlations in large networks. We conclude that rank correlations are a suitable {and} informative method for uncovering network mixing patterns. 

For the correct interpretation of degree-degree dependencies, it is important to realise that positive or negative correlations can be pre-defined by the degree sequence itself. For instance, there is only one simple graphs with degrees $(3,1,1,1)$, and the result $\rho_4=-1$ is not informative in this case. It has been discussed in the literature that, conditioned on not having self-loops and multiple edges, random networks with given degrees exhibit disassortative patterns~\cite{Boguna2004finite-size-effects,Maslov2004rewiring,Park2003degree-correlations}, also called {\it structural} correlations. In order to filter out the structural correlations, one needs to compare the real-world networks to their null-models -- graphs with the same degree sequences but \emph{random} connections. This null-model is a uniform simple random graph with the same degree sequence. Here a network is called simple when it has no self-loops nor multiple edges. Such a graph can be obtained by randomly pairing half-edges, as in Section~\ref{sec:RG}, and taking the first realization that is simple. This is especially problematic when $(\max_i d_i)^2>|E|$, which is the case in many examples, since then one needs a prohibitingly large number of attempts before a simple graph is generated~\cite{Janson2009CM, HofstadRG}. 

A widely accepted method for constructing a null-model, is the random rewiring of the connections in a given graph~\cite{Maslov2002rewiring,Maslov2004rewiring}. The disadvantage is the unknown running time before a graph is produced that is close enough to being uniform. Recent work \cite{Blitzstein2011conf_model} presents a sequential algorithm, where, at each step, the remaining unconnected edges maintain the ability to generate a simple graph. This method always produces the desired outcome but its worst-case running time $O(n^2\sum_i d_i)$ is infeasible for large networks. The recently introduced grand-canonical model~\cite{Squartini2011MLM} computes the probability of connection between two nodes in a maximum entropy graph with given degree sequence, and enables the evaluation of many characteristics of the graph. To the best of our knowledge, efficient implementation of this method for large networks has not been developed yet. 

Constructing a null-model and filtering out the structural correlations in large networks is an interesting and demanding computational task that is beyond the scope of this paper. We believe that structural correlations will affect $\rho_n$ to a larger extent than the rank correlation $\rho_n^{\rm rank}$ because it is usually the nodes with largest degrees that produce self-loops and multiple edges, and thus the relative contribution of these edges in the cross-products will be larger for $\rho_n$ than for $\rho_n^{\rm rank}$. This conjecture requires a further investigation. 

We conclude by stating that rank correlation measures deserve to become a standard tool in the analysis of complex networks. The use of rank correlation measures has become common ground in the area of statistics for analysing heavy-tailed data. We hope to have provided a sufficient evidence that this method is preferred for analysing network data with heavy-tailed degrees as well.}


\section*{Acknowledgment}

We thank Yana Volkovich for the code generating a Preferential Attachment graph.
{This article is also the result of joint research in the 3TU
Centre of Competence NIRICT (Netherlands Institute for Research on
ICT) within the Federation of Three Universities of Technology in
The Netherlands. The work of RvdH was supported in part by the Netherlands Organisation for Scientific Research (NWO). The work of NL is partially supported by the EU-FET Open grant NADINE (288956).}

\bibliographystyle{apsrev4-1}
\bibliography{myrefs}

\end{document}